\documentclass[11pt,a4paper]{article}
\usepackage{amssymb} 

\def \be {\begin{equation}}
\def \ee {\end{equation}}
\def \bea {\begin{eqnarray}}
\def \eea {\end{eqnarray}}
\def \pp {{\Bbb P}}
\def \zz {{\Bbb Z}}
\def \cc {{\Bbb C}}
\def \oo {{\cal O}}
\def \pbp {Dp-D \bar p}

\begin{document}
\begin{titlepage}
\begin{flushright}
SISSA/92/99/FM\\
hep-th/9908009\\
\end{flushright}

\begin{center}
{\Large\bf
Projective resolutions of coherent sheaves\\
and descent relations between branes.\\}

\vspace{.5cm}
{\small\bf A.Tomasiello}

\vspace{.5cm}

{\small\sl  
International School for Advanced Studies \\
via Beirut 2-4, 34014 Trieste, Italy\\
E-mail: {\tt tomasiel@sissa.it}}
\end{center}

\begin{abstract}
\noindent
We notice that, for branes wrapped on complex analytic subvarieties, 
the algebraic-geometric version of K-theory makes the
identification between brane-antibrane pairs and lower-dimensional branes 
automatic. This is because 
coherent sheaves on the ambient variety represent gauge
bundles on subvarieties, and they can be put in exact sequences 
(projective resolutions) with sheaves
corresponding to vector bundles on the pair; this automatically
gives a $D(p-2)$ as a formal difference of bundles on the $\pbp$ 
pair, both belonging to
 the Grothendieck group of coherent sheaves of the ambient. 
\end{abstract}

\vfill
\end{titlepage}

\section{Introduction.}

There has been recently a major shift of perspective as to the interpretation
of tachyons in brane-antibrane pairs \cite{sen}. It has been long since 
suspected that they
signal instabilities, but one could not forecast whether this led to
inconsistency of the theory or could have instead a more constructive role.
It is now believed that the latter is the case, and that tachyons can be
viewed simply as Higgs fields, whose condensation can lead to interesting
effects. In particular, it can lead to remnants whose stability is in turn 
simply dictated by the topological stability of the condensation. Here we will
be interested only in type II theories and in stable products of decay. 
If on the brane and on the antibrane there are gauge bundles $E$ and $F$
respectively, what we need is something that classifies the
topological content of the pair. The key is to note that \cite{witten}, 
if the two bundles
are topologically the same, they will annihilate leaving no stable remnant:
there is in this case no stable pattern of condensation for the tachyon.
Moreover, if such a trivial pair is added to non trivial ones, it will
disappear, giving no contribution. We can write this symbolically
as $ (E,F) \sim (E\oplus H , F\oplus H)$. Since this is the definition of
topological K-theory, we can interpret physics by identifying a brane-antibrane
system as an element of this group (hints in this direction were explicitly 
given before in \cite{minmoore}).

We have now to identify physically the result of the decay, which in general 
is not the vacuum (as there exist non-trivial elements of K-theory).
The locus in which the tachyon vanishes is the place around which energy is
located, and is a source for lower RR fields. So it is natural to identify it
with a brane supported on its zero locus;
this can be viewed as a section of ${\cal T} \equiv E\otimes F^*$. This second
point makes it possible that K-theory also classifies all lower-dimensional
D-branes.

Let us stress that the classification of pairs by K-theory and the
identification of the remnant are, so far, two different steps. 
Note, indeed, 
that a K-theory class does not define
the bundle $\cal T$: if we change the representative, the latter
changes as well.

Here we will argue that, if we deal with complex analytic subvarieties,
the identification of the remnant is a natural consequence of the first step,
in a way that puts a new light on the incorporation of lower charges.
This is because in that case we can choose, instead of K-theory, an analytic
(algebraic-geometric) version of it, the Grothendieck group of coherent
sheaves \cite{hart, manin}, 
hereafter referred to simply as Grothendieck group (as it is commonly meant
in algebraic geometry; in general this
name simply denotes the procedure to get a group from a monoid).
In this setting, bundles over subvarieties are expressible - through projective
resolutions - as formal differences of bundles over the ambient variety.
This allows us to see 
the identification between brane-antibrane and K-theory on the same footing 
as that between 
brane-antibrane and lower-dimensional brane; in a way, this becomes 
automatic, or natural. 

On the one hand, this seems to confirm the physical analysis already done; on
the other, it allows us 
to interpret descent relations, and tachyon condensation
between stable D-branes, in
terms of projective resolutions. In turn, the physical identification between
$\pbp$ and $D(p-2)$ can be viewed now as a purely topological counterpart of
an algebraic-geometric procedure.

One should stress the difference between what we said above and the
already known mathematical counterpart of tachyon condensation: the
isomorphism between the K-theory of a submanifold and that of the ambient. 
While that is an isomorphism between different groups, what we propose here is
that the descent relation can be viewed as natural in one {\sl single} group.
Even better, while that isomorphism is
natural, it does not justify in itself the identification of the $D(p-2)$ with
the pair. 
In our case, on the contrary, we treat complex analytic subvarieties and
so we can choose to use Grothendieck groups: if we do that, the identification
of the product of tachyon condensation is {\sl a logical consequence} of the
existence of projective resolutions of coherent sheaves.
A related issue is the following.
One often says that K-theory of the ambient space classifies branes: if so,
not only should there be representatives in the group for all
lower-dimensional branes, but also different classes for branes of different 
dimensions. While the first feature has been demonstrated to be there, the
second would lack if we did not choose different boundary conditions (and
thus compactifications of the ambient space, by adding points at infinity; for
instance usually one takes spheres to classify branes on flat space), 
for example by considering higher
K-theory groups. These groups together then form a ring which one considers
as classifying branes. Our attitude here will be different: we will have 
everything in
the same group, by taking, as compactification, complex projective space, which
has the virtue of giving us a $\zz$ for every codimension. 
We will return to this later.

The following discussion can be fitted to any case in which we want to
express a brane wrapped on a submanifold in terms of a pair wrapped on a
manifold; just compactify the latter to a complex projective variety.  
The main case we have in mind is IIB theory, with space filling
branes and antibranes; so in this case we take, as we said, 
projective space $\pp^5$, of real dimension $10$, as ambient variety.
The examples that we will be talking about in the
following, however, are chosen to be easier. 
We do not make any claim of mathematical originality; these
examples are here just to clarify and illustrate 
the general discussion, which also
relies on well-known mathematical facts.

\vspace{.3cm}
Issues about the role of Grothendieck groups were already considered in 
\cite{sharpe}, with a different point of view and aim from the one 
we take here.

\section{Coherent sheaves and their projective resolutions.}
The key concept we have to introduce, before discussing
Grothendieck groups, is that of {\it coherent sheaf}.
Sheaves generalize the concept 
of fibre bundles on a manifold, which can be viewed as a particular class 
of sheaves, the locally free ones. Fibre bundles on a submanifold will
then be described precisely by coherent
sheaves. On a projective smooth variety $X$, the case to which for 
simplicity we will now
restrict, they are sheaves of finitely generated ${\cal O}_X$ -modules, and
thus (using, for example, 
regularity and the presence of an ample invertible sheaf) 
admit a 
finite projective free resolution, that is, 
they can be put in an exact sequence:
\be
0 \to {\cal F}_{n-1} \to \ldots \to {\cal F}_1 \to {\cal F}_0 \to {\cal C} 
\to 0
\label{defseq}
\ee
where $\cal C$ is our coherent sheaf and the ${\cal F}_{k}$ are locally free.
Fibre bundles on a closed subvariety are coherent sheaves 
(closed immersions are
proper, and so take coherent sheaves in the subvariety -- in particular,
locally free ones -- to coherent sheaves), and so admit such resolutions; 
we will see how this works in examples. Note that our conclusions are
untouched by possible singularities of the subvarieties.

There are in general two definitions of Grothendieck groups. 
Take $\zz[{\cal L}]$ 
(respectively 
$\zz[{\cal C}]$) the free abelian group generated by locally free
(respectively coherent) sheaves. Then 
$$
{\cal K}^{\bullet} \equiv \zz[\cal L] / {\cal I}, \qquad 
{\cal K}_{\bullet} \equiv \zz[\cal C] / {\cal I'}, 
$$
where  $\cal I$ ($ \cal I '$) is the ideal generated in both cases by elements
of the form ${\cal F} - {\cal F}' - {\cal F}'' $ if there is an exact sequence
$ 0 \to {\cal F}'' \to {\cal F} \to {\cal F}' \to 0. $
In the case of smooth projective varieties, these two definitions are the
same (the two groups are isomorphic). In particular we can immediately see 
that every coherent sheaf can be
expressed as an element of ${\cal K}^{\bullet}$, by their defining sequence
(\ref{defseq}), 
and so as the difference 
$\cal E - \cal F$ of two locally free sheaves.
This simply means that in this setting we can express a fibre bundle on a
subvariety as the difference of two fibre bundles on the variety. This is just
the identification given by tachyon condensation! So we see that, if we
use Grothendieck group instead of topological K-theory, the first step we
described above (the fact that we can cancel the same bundle if it appears on
both sides) implies the second (tachyon condensation and identification with
lower-dimensional branes). Of course we have not shown the isomorphism between
Grothendieck groups here: but this point is also essential,
because otherwise the class which we found would
depend on the projective resolution we chose. 
Although we will not prove that here
\cite{manin}, we will check it explicitly in an example at the end of the
paper.

This group ${\cal K}^{\bullet}\cong {\cal K}_{\bullet}$
is different from K-theory in various respects. First, of course,
we are working with holomorphic subvarieties, and choosing holomorphic
structures for the bundles involved. Second, what we identify with zero are
extensions instead of direct sums. But all extensions are topologically
trivial (direct sums), so this is not a problem. 
So this group does not contain in general all of what is contained in K-theory
(the non-complex bundles); this gives us a means to understand which
brane-antibrane configurations represent lower dimensional branes wrapped on
holomorphic cycles. At the same time this group distinguishes objects which in
K-theory are the same.

We have still to verify that the isomorphism between the Grothendieck groups 
is really consistent with the tachyon condensation we already know. We will do
this in several examples, using as a basic tool the sequence
\be
0 \to {\cal F} \to {\cal F}(D) \to {\cal F}(D)_{|_D} \to 0,
\label{fund}
\ee
with $\cal F$ locally free (alternatively, we can derive it from 
(\ref{oo}) tensoring).
Start from a brane wrapped on a holomorphic cycle with its bundle - that is, a
coherent sheaf. We have to express it in terms of locally free sheaves
(brane-antibrane) and then do tachyon condensation to see whether we retrieve
the initial configuration.
Let us start with the case of one brane, with the structure sheaf as bundle on
it (we think of the $U(1)$-bundle as the circle bundle inside it), and 
suppose that we work in $X=\pp^n$. The resolution depends on
whether the subvariety $Y$ is a 
complete intersection or not; in the former case
it is easy. In the case of (real) codimension 2 it reads
\be
0 \to {\cal O}_X (- Y) \to {\cal O}_X \to {\cal O}_Y \to 0;
\label{oo}\ee
the first term is the ideal sheaf, that we can think as 
the sheaf of holomorphic functions having a zero along $Y$.
From this we get that 
${\cal O}_Y= {\cal O}_X - {\cal O}_X( -Y)$,
and so the tachyon is a section of 
${\cal E} \otimes {\cal F}^* = {\cal O}_X (Y)$,
 hence it has a simple
zero on $Y$ (recall there is a duality between functions having poles and
sections having zeroes). This exactly parallels the already known 
construction of a brane-antibrane pair which reproduces a codimension 2 brane. 
We already remarked that the bundle of which the tachyon is
a section is not defined by the class in K-theory; instead, it is guessed in
some way by thinking about a ``minimal'' couple $(E,F)$. With this procedure,
this minimality is just the concept of minimal resolution of a coherent sheaf,
well-known to algebraic geometers.

\vspace{.3cm}
\leftline{\underline{\sl{Higher codimension.}}}
\vspace{.2cm}
Let us describe what to do if the codimension is higher, sticking with the case
of complex projective spaces. Exploiting the fact that line bundles, in this 
case, are
classified by $\zz$, we only write degrees. Let $d_1,\ldots, d_l$ be the
degrees of the equations $f_1, \ldots, f_l$ defining the subvariety, and  
suppose it to be a complete intersection,
\bea
0 \to {\cal O}_X (-d_1 - \ldots -d_l) \to \ldots & \to & 
\bigoplus_{i_1, i_2} {\cal O}_X ( - d_{i_1} - d_{i_2}) \label{syzygy}   \\
\nonumber &\to &
\bigoplus_{i} {\cal O}_X ( -d_i) \buildrel {\phi}\over\to
 {\cal O}_X \to {\cal O}_Y \to 0.
\eea  
The origin of this resolution can be understood as follows: the 
$ {\cal V} \equiv\oplus_i \oo(-d_i)$ term 
is the generalization of the first (from the left) term of 
(\ref{oo}) due to the fact that we have more than one equation, and may be
thought as the sheaf of $l$-uples of functions $(f_1 h_1, \ldots, f_l h_l)$;
$\phi$ takes such an $l$-uple 
to $\sum f_i h_i$, and the next term on the left is its kernel,
i.e. relations between the $f_i$.
We always find the obvious relations $f_i \cdot f_j - f_j \cdot f_i=0$,  
which generate the term $\oplus_{i,j} \oo(-d_1 -d_j)$; in the case of non
complete intersections, the number of equations necessary to define 
the variety 
would be higher than the codimension, and we would have extra terms. 
In fact in the complete intersection case one may make (\ref{syzygy}) more 
transparent in the form 
$$ 0 \to \Lambda^l {\cal V} \to \ldots \to \Lambda^2 {\cal V} 
\to \Lambda {\cal V} \to \oo_X \to \oo_Y \to 0, $$ 
also known as the {\it Koszul complex}. 
Note that we do not have to proceed step by step: this is an ``all at once''
procedure, 
as the one already known and involving K-theory Thom isomorphism and 
ABS construction
(\cite{witten}; see also \cite{szol} for a more extensive description).
To see better the correspondence, first remember that for any complex 
$$ E_0 \buildrel {d_0}\over\to E_1 \buildrel {d_1}\over\to  \ldots 
\buildrel {d_{n-1}} \over \to E_n$$ 
one may construct a smaller complex (a map) 
$$ \bigoplus_{i=\rm{even}} E_i\buildrel {D}\over\to
\bigoplus_{i=\rm{odd}} E_i,$$
where $D(e_0, e_2, \ldots) = (d_0 e_0 + d_1^* e_2, d_2 e_2 + d_3^* e_4,
\ldots)$; moreover the latter complex is exact if the former is. 
If we apply this to the sequence (\ref{syzygy}), then we have a 
pair of bundles which just correspond to the Clifford bundles $({\cal S_+,
S_-})$ associated to the conormal of our variety; for example 
Clifford multiplication (which can be viewed as a sum of an exterior product
and a contraction) arises from the sum in the definition of $D$. 

Projective resolutions like this are also known, when thought in terms of the
corresponding algebraic objects (rings and modules) as 
{\it syzygies} \cite{harris}.

\vspace{.3cm}
\leftline{\underline{\sl{Lower charges.}}}
\vspace{.2cm}

We will now give more concrete examples for the case of a non trivial 
line bundle on the $(p-2)$-brane. Since we will do this again in projective
spaces, let us return for a moment to the point we anticipated 
in the introduction.
One would imagine, from the
descriptions we gave so far of the identifications of the $\pbp$
system with lower dimensional 
branes, that, fixed our ambient space, K-theory or its Grothendieck group
already provides a classification of all branes of all dimensions. To
check this we have to choose the space; even if we have fixed 
the dimension $p$, 
we still have to choose the boundary conditions. 
This amounts to choosing a type of compactification. We can
choose, as usual, a sphere; but the K-theory of a sphere is at most $\zz$. 
This means that the construction, by itself, would confuse branes of
different dimensions: two dimensionally different branes can be realized as
the same $\pbp$ system. We have, so to say, to avoid this by hand,
treating each codimension separately and imposing that our bundles depend just
on the transversal dimensions. If we do not do that, we find that in a given 
brane-antibrane system, a $(p-2)$ brane is given by a certain couple $(E, F)$,
which is anyway also topologically the same as a couple which gives a $(p-4)$
brane.

We point out that it could be that different choices of boundary conditions
(that is, of compactification) can give a better result. For instance, choosing
the complex projective space $\pp^n$, both K-theory
and the Grothendieck group are isomorphic to \cite{manin,atiyah}
\be
R \equiv \zz [t] / (t-1)^{n+1},\label{pn}
\ee
which gives a $\zz$ for each even codimensional brane
(the isomorphism from ${\cal K}_{\bullet}$ or K-theory  to $R$ is given by 
sending the hyperplane $H$ to $t$). The fact that the topological and
algebraic geometrical K-theory are equal here is reasonable, since we know
that $\pp^n$ has no Jacobian. In a way the trick is that, choosing $\pp^n$ as
compactification space, we allowed all possible boundary conditions, so
distinguishing between branes of different dimensions. Of course 
in this way we are not predicting the allowed codimensions of the branes: we
cannot because we chose the complex analytic setting, forcing them to be even.

Now we return to the examples we wanted to give: non trivial line bundles on
the $(p-2)$ brane. This is easier if we can extend
the line bundle to one on the ambient space, whose divisor let us call $D$: 
indeed, applying again (\ref{fund}), we obtain
$$
0 \to {\cal O}_X (D- Y) \to {\cal O}_X (D) \to {\cal O}(D)_{|_Y} \to 0;
$$
the last term is just our
original bundle on the subvariety. Note that again 
${\cal E} \otimes {\cal F}^* = {\cal O}_X (Y)$, as it should.
To fix ideas, consider the case of a conic in $\pp^2$, which is
birationally equivalent to $\pp^1$. If the line bundle is 
${\cal O}_{\pp^1}(2)$, 
it can be extended on $\pp^2$ as
${\cal O}(1)$ (henceforth $\oo \equiv \oo_{\pp^2}$); the sequence so reads
$$
0 \to {\cal O}(-1) \to {\cal O}(1) \to 
{\cal O}_{\pp^1}(2) \to 0.
$$

Life is harder if we consider instead the bundle 
${\cal O}_{\pp^1}(1)$, which is not extendable to $\pp^2$, since there is no
divisor on it which intersects the conic once. We can circumvent this
difficulty in two ways. First, we can start the projective resolution 
from $\oo \oplus \oo$, instead
of from $\oo$. Doing this carefully involves looking
at the rings of functions over the varieties, and maps between them; we will 
not
delve into details, but the result turns 
out to be 
$$ 0 \to 2 \oo(-1) \to 2 \oo \to \oo_{\pp^1}(1)\to 0$$
(we denote now $\oplus_1^n \oo(k)$ as $n \oo(k)$).
As a check, let us do it in another way, which only involves sequences that we
have already seen. Write first the two sequences
\be 
0 \to \oo_{\pp^1}(-1) \to \oo_{\pp^1} \to \oo_p \to 0
\label{pt}
\ee
$$ 0 \to \oo(-2) \to 2 \oo(-1) \to \oo \to \oo_p \to 0$$  
where $p$ is a point in $\pp^1$ at which a section of our line bundle
$\oo_{\pp^1}(1)$ vanish. Now, tensor both with an hyperplane line bundle in
$\pp^2$: we get  
$$
\oo_{\pp^1}(1) = \oo_{\pp^1}(2) - \oo_p(1)= \oo(1) - \oo(-1) -  
(\oo(1) -2 \oo + \oo(-1))= 2 \oo - 2 \oo(-1)
$$
as above. Still another method would have been tensoring the first sequence in
(\ref{pt}) by an hyperplane line bundle of $\pp^1$, and then reexpressing 
 $\oo(1)$ as 
$3\oo -3 \oo(-1) + \oo(-2)$: this gives an illustration of the relation
$(t-1)^{n+1}=0$ in (\ref{pn}). The result is in any case reasonable because a
conic can be deformed to two hyperplanes touching in a point.

If finally we want to consider stacked branes, $U(1)$-bundles become 
$U(r)$-bundles. So the generalization is to represent $U(r)$ on $\cc^r$; we
get in this way a complex vector bundle of rank $r$ on the
lower-dimensional brane. The coherent sheaf is now again a sheaf of ${\cal
O}$-modules, but with more generators; and so the resolution starts with 
$r\oo_X$ instead of with $\oo_X$. This is right, since to get
stacked branes we expect to have to start from stacked brane-antibrane pair.
The procedure works the same way, but to verify that we get the desired
subvariety we have to look at the locus where the tachyon is non invertible,
rather than zero.

\vspace{.5cm}
We thank L.~Bonora, A.~Sen, A.~Zampa for useful discussions and advice.

\end{document}